\documentclass[11pt, a4paper]{article}

\usepackage{amsmath}
\usepackage{amsfonts}
\usepackage{epstopdf}
\usepackage{epsfig}
\usepackage{latexsym}
\usepackage{color}
\usepackage{amsmath,bm,amssymb}
\usepackage{cite}
\usepackage{slashed}
\usepackage{soul}
\usepackage[hyperfootnotes=true]{hyperref}
\hypersetup{colorlinks, citecolor=black, linkcolor=black, urlcolor=black}
\definecolor{bluscuro}{rgb}{0.15, 0.2, .85}
\usepackage[multiple]{footmisc}

\newcommand*{\LargerCdot}{\raisebox{-0.25ex}{\scalebox{1.2}{$\cdot$}}}

\newcommand{\eq}[1]{(\ref{#1})}

\newcommand{\be}{\begin{equation} }
\newcommand{\ee}{\end{equation} }
\newcommand{\bs}{\begin{split} }
\newcommand{\es}{\end{split} }

\usepackage{accents}

\definecolor{GRed}{rgb}{0.9,0.12,0.1}

\newcommand{\J}{\mathcal{J}}

\newcommand{\ch}{\mathcal{H}}

\begin{document}

\begin{titlepage}

\vspace{1cm}
\begin{center}
{\huge
The effective theory of fluids at NLO and \\implications for dark energy}
\\
\bigskip\color{black}
\vspace{1cm}{
{\large Guillermo Ballesteros}
\vspace{0.3cm}
} \\[7mm]
\emph{Institut f\"{u}r Theoretische Physik, Universit\"{a}t Heidelberg\\ Philosophenweg 16, D-69120 Heidelberg, Germany.\\\vspace{0.2cm} Instituto de F\'{i}sica Te\'{o}rica, IFT-UAM/CSIC\\ Cantoblanco 28049, Madrid, Spain.}\vspace{0.3cm}

{\small guillermo.ballesteros@uam.es}

\end{center}
\bigskip

\vspace{.4cm}

\begin{abstract}
We present the effective theory of fluids at next-to-leading order in derivatives, including an operator that has not been considered until now. The power-counting scheme and its connection with the propagation of phonon and  metric fluctuations are emphasized. In a perturbed FLRW geometry the theory presents a set of features that make it very rich for modelling the acceleration of the Universe. These include anisotropic stress, a non-adiabatic speed of sound and modifications to the standard equations of vector and tensor modes. These effects are determined by an energy scale which controls the size of the high derivative terms and ensures that no instabilities appear.
\end{abstract}

\end{titlepage}

\section{Introduction}

One of the major problems of cosmology is understanding the origin of the observed acceleration of the Universe \cite{Riess:1998cb,Perlmutter:1998np}. A cosmological constant is sufficient to fit the current data,  but the difficulty to explain its tiny measured value \cite{Weinberg:1988cp} has led to a large amount of alternative models. Although none of these models really solves the cosmological constant problem, it is nevertheless interesting to consider the best motivated among them. This helps to devise strategies to rule out the standard $\Lambda$CDM scenario by searching for departures from it using various types of observations such as Type Ia supernovae, baryon acoustic oscillations, weak gravitational lensing and galaxy clusters (see e.g. \cite{Weinberg:2012es} for details). New satellite (Euclid \cite{Laureijs:2011gra,Amendola:2012ys}, WFIRST \cite{Green:2012mj}) and ground based (DESI, see e.g.\ \cite{Levi:2013gra}, and LSST \cite{Abate:2012za}) probes specifically designed for this purpose will start operating in the coming years.

A large class of models of the acceleration is based on a scalar field whose time evolution allows to mimic the knowledge we have about the expansion history of the Universe. These models have been steadily increasing in complexity from the first proposals \cite{Wetterich:1987fm, Ratra:1987rm}.  In particular, much interest has been devoted recently  to the study of large subclasses of models of this type with second order equations of motion \cite{Horndeski:1974,Deffayet:2009wt,Deffayet:2011gz,Zumalacarregui:2013pma,Gleyzes:2014dya,Gao:2014soa}, in order to avoid potential ghost or gradient instabilities, see e.g.\ \cite{Woodard:2006nt}. 

Although such models are certainly interesting, it may well be that if anything beyond a cosmological constant drives the acceleration it might not actually be a fundamental scalar. It is not unreasonable to think that many of these models should rather be regarded as a practical {\it effective} description of the acceleration. In fact, if the acceleration were the low energy consequence of a yet unknown physical phenomenon occurring at higher energies, the machinery of effective field theories should be used to analyze it. However, many of the scalar field models that have been proposed are not well-defined effective theories, e.g. \cite{Horndeski:1974}. 

Considering as an example General Relativity, we may assume that the acceleration is due to some sort of substance, dark energy, that fills the universe feeding the right hand side of Einstein's equations $G_{\mu\nu}=8\pi G\, T_{\mu\nu}$. On very general grounds, this dark energy may be thought of as a coarse-grained continuum (or fluid) that acts on the curvature of spacetime. If so, it should be possible to characterize its elements, at least partially, using three coordinates that indicate their positions in space at any given time. By promoting these coordinates to spacetime fields, it turns out that this simple picture allows to build a powerful effective field theory framework that we can apply to describe the acceleration of the Universe. This construction is as close to first principles as possible and is entirely determined by the symmetries of the continuum.

Fluids can be described with an effective field theory in which the set of relevant low-energy degrees of freedom contains (but is not necessarily limited to) scalar fields that represent the comoving coordinates of the fluid itself \cite{Dubovsky:2005xd}. In this picture, the propagation of gapless sound waves (phonons) in the fluid is given by the dynamics of Goldstone bosons that appear as a consequence of the spontaneous breaking of the internal symmetry that characterizes the properties of the fluid. In a simple case, this symmetry is the invariance under the group $\mathcal{V}$Diff of (internal) diffeomorphisms that preserve the volume. The corresponding effective action at lowest order gives a perfect fluid \cite{Dubovsky:2005xd}. In three spatial dimensions there are three Goldstone bosons, that can be separated in two (transverse) divergenceless fields and a (longitudinal) irrotational one. In general, the effective theory is organized as a derivative expansion and if the symmetry is $\mathcal{V}$Diff the leading order is determined by a single operator of the comoving coordinates. 

This framework (see \cite{Leutwyler:1996er,Leutwyler:1993gf} for earlier work) is very rich and has been used to describe a wide variety of systems including  e.g.\ non-dissipative fluids, supersymmetric fluids \cite{Hoyos:2012dh} and supersolids \cite{Son:2005ak}, among other possibilities. The minimal $\mathcal{V}$Diff example can be modified by either changing the symmetry or enhancing the field content \cite{Dubovsky:2005xd,Dubovsky:2011sj}. For instance, lowering the symmetry requirement from $\mathcal{V}$Diff to internal translations and $SO(3)$ rotations (which corresponds to a solid) modifies qualitatively the dynamics of transverse phonons. Whereas in the $\mathcal{V}$Diff case the transverse modes behave in Minkowski as standing waves, they do propagate once the symmetry is relaxed. 

Several applications of effective fluids in cosmology already exist. Perturbations of an effective perfect fluid in FLRW at leading order in derivatives were studied in \cite{Ballesteros:2012kv}. A model of inflation with novel non-Gaussian signatures based on the symmetry properties of a solid was proposed in \cite{Gruzinov:2004ty, Endlich:2012pz}. The issue of defining different cosmological species and their interactions in multi-component fluids was dealt with in \cite{Ballesteros:2013nwa}. The effective theory of fluids is related to the pull-back formalism (see \cite{Andersson:2006nr} for a review), which has also been applied in cosmology, e.g.\ in \cite{Blas:2012vn,Pourtsidou:2013nha}.

Most works on effective fluids have been limited to studying the theory at the lowest order in derivatives, but there are some that have examined the properties of certain operators beyond leading order \cite{Dubovsky:2011sj,Nicolis:2011ey,Bhattacharya:2012zx}. Since fluids are described by a derivatively coupled effective theory, they generically have equations of motion with high order derivatives. This is of no concern as far as the theory is applied at energies sufficiently lower than its cut-off. Therefore, there is no need at all  of building a finely tuned Lagrangian to avoid the appearance of higher order derivatives in this context. By construction, the operators with high derivatives are suppressed and hence the theory protects itself from developing instabilities. This is in striking contrast to the approach mentioned before and that is so often advocated for popular models of a single scalar field, e.g.\ \cite{Horndeski:1974}.

The main purpose of this paper is presenting the effective field theory of fluids at next-to-leading order in derivatives and showing (with a simple symmetry choice) the rich phenomenology that appears when it is applied for the description of the acceleration of the Universe. 

The outline of the paper is the following. In Section \ref{deft} we review the effective theory of fluids as a particular example of a broad class of theories for derivatively coupled scalars. In Section \ref{eftph} we focus on the dynamics of phonons in the effective framework, emphasizing the power-counting scheme, listing the relevant operators at next-to-leading order in derivatives and deriving the equations of motion for transverse and longitudinal modes in Minkowski spacetime. In Section \ref{cosmos} we write the action at second order in derivatives for an effective fluid including gravity and study its properties in a perturbed FLRW universe. The conclusions are presented in Section \ref{conc}. 

\section{Fluid from comoving coordinates} \label{deft}
In this section we review the construction of the effective theory of fluids following mainly \cite{Dubovsky:2005xd}, where it is shown that it is a particular example of a broader class of derivatively coupled effective field theories. Let $\{\phi^I\}$ be a set of indexed real scalar fields of dimension 1 (in units of energy) and write the action \cite{Dubovsky:2005xd} 
\begin{align} \label{gena}
\Lambda^4\int d^4x\, \sqrt{-g}\, F\left(\epsilon \frac{\phi^I}{\Lambda},\frac{B^{IJ}}{\Lambda^4},\frac{\Box\phi^I \Box\phi^J}{\Lambda^6},\ldots\right)\,,
\end{align}
where we define 
\begin{align} \label{Bmatrix}
B^{IJ}=g^{\mu\nu}\partial_\mu\phi^I\partial_\nu\phi^J
\end{align} 
and use the notation $\Box =\nabla_\mu\nabla^\mu$, being $\nabla_\mu$ the covariant derivative associated to the metric $g_{\mu\nu}$.
The action \eq{gena} depends explicitly on a constant $\epsilon\ll 1$ and an energy scale $\Lambda\ll M_P=1/\sqrt{8\pi G}$, where $G$ is Newton's gravitational constant. Any other (dimensionless) parameters contained inside the dimensionless function $F$ are assumed to be comparable to 1. More generally, $F$ and all its derivatives with respect to its arguments are of order 1. In the action \eq{gena} it is assumed that given two operators with the same number of fields, the one with less derivatives is more important. Schematically, we can write this as  $\partial \ll \Lambda$ where $\partial$ represents any first order spacetime derivative. More explicitly, assuming $|\partial\phi|\sim|\partial\phi|^q$ for any real $q$, we can write
\begin{align} \label{pcphi}
\Lambda^{m}|\partial\phi|^n\ll|\partial^m|\partial\phi|^n|\,.
\end{align}
This defines a derivative power-counting scheme for the action \eq{gena}. With the previous assumptions, the effective mass matrix is $ m_{IJ}\,^2 \sim\epsilon^2\Lambda^2$. In the limit of vanishing $\epsilon$, the fields $\phi$ become massless and the theory only has derivative couplings. As discussed in \cite{Dubovsky:2005xd}, if we add an Einstein-Hilbert piece to \eq{gena}, the conditions on $F$ constrain the curvature $R$ to be of the order of  ${\Lambda^4}/{M_P^2}$, because $M_P^2 G_{\mu\nu}=T_{\mu\nu}$ implies that $M_P^2 R\sim \Lambda^4$. This means that the spacetime is curved by the fields $\Phi$ on distance scales of the size of
${M_P}/{\Lambda^2}\gg {1}/{\Lambda}$. Then, there is a range of momentum scales:
$ \max\left(\Lambda^2/M_P\,,\epsilon\Lambda\right) \ll k \ll \Lambda$, where the $\phi^I$ fields are approximately massless and the spacetime can be described by Minkowski. In this range, it is possible to approximate the $\phi^I$ at first order  by affine functions of the spacetime coordinates and the action \eq{gena} collapses at lowest order in derivatives to:
\begin{align} \label{genaction}
\Lambda^4 \int d^4x\, F\left({B^{IJ}}/{\Lambda^4}\right)\,.
\end{align}

\subsection{The effective theory of perfect fluids} \label{secfluids}
The (lowest order) effective field theory of (perfect) fluids in Minkowski spacetime is a subclass of the action \eq{genaction}. To see this we are going to work with three scalar fields, $\Phi^a$, analogous to the fields $\phi^I$ of the previous section. These three scalars $\Phi^a$ can be interpreted as the comoving coordinates of the continuous medium (the fluid) that we intend to describe with an effective theory. It is this picture that renders the theory that we are about to present a simple description of continuous media in an effective framework. As we will see, the physical properties of the fluid are characterized by the symmetries imposed on the action for the fields $\Phi^a$. 

So, let us consider three spacetime scalar functions $\Phi^{a}\in \mathbb{R}$ with $a\in\{1,2,3\}$\,, that allow to define, at any fixed time, an isomorphism between the three-dimensional coordinate space of an observer $\mathcal{O}$ and a continuum of points $\mathcal{F}$ (that can be thought to represent the fluid). These functions $\Phi^a$ label the points (elements of the continuum) in $\mathcal{F}$, whereas the map $\Phi\longrightarrow x(\tau,\Phi)$ gives the position in real space of the fluid element $\Phi$ at a given time $\tau$.
In other words, $x^i=x^i(\tau,\Phi)$ is the trajectory of the element $\Phi$.  The inverse map, $x\longrightarrow \Phi(\tau,x)$, returns the fluid element $\Phi$ that is sitting at the position $x$ at the time $\tau$.  At any time, the spaces $\mathcal{O}$ and $\mathcal{F}$ are isomorphic and the isomorphism can be chosen such that $\Phi^a(\tau,x^i)=x^a$ in a stationary state. The Lagrangian for $\Phi$ satisfying that the stationary state
\begin{align} \label{ground}
\Phi(\tau,x)=x
\end{align}
is invariant under a combination of a space translation $x \rightarrow x-c$ and a internal translation (or shift symmetry)
\begin{align} \label{shift}
\Phi\rightarrow \Phi+c 
\end{align}
has to be necessarily an arbitrary function of \eq{Bmatrix} at the lowest derivative level, precisely as in eq.\ \eq{genaction}. Notice that if the $\Phi$ fields are interpreted as the comoving coordinates of the fluid, it is convenient to assign to them dimensions of length, in which case the matrix $B$ defined in \eq{Bmatrix} is dimensionless. We will follow this convention to avoid introducing a dimensionful scale in the definition of the ground state \eq{ground}. To simplify further the notation, in the following we will absorb the multiplicative factor $\Lambda$ of the action \eq{genaction} into the definition of the function $F$, which will then have dimension four. 

The symmetry \eq{shift} can be enhanced to the invariance under volume-preserving internal diffeomorphisms $\mathcal{V}$Diff:
\begin{align}
 \label{diffs}
 \Phi^a\longrightarrow  \Psi^a(\Phi)\,, \qquad  \det\left(\frac{\partial \Psi^a}{\partial \Phi^b}\right)=1\,,
 \end{align}
in which case the action \eq{genaction} becomes
\begin{align} \label{sfluid}
\int d^4x\, F(b)\,,
\end{align}
where
\begin{align} \label{detb}
b=\sqrt{\det B}\,,
\end{align}
and $F$ is a real-valued function with dimensions of energy density. As we will later see in more detail, this expression is actually the most general action at lowest order in the derivative expansion of the perturbations of $\Phi$ around the background \eq{ground}. At higher orders, the function $F$ should include other invariants in addition to $b$.

The action \eq{sfluid} can be generalized to arbitrary spacetimes through direct covariantization:
\begin{align} \label{sflu}
\int d^4x\,\sqrt{-g}\, F(b)\,.
\end{align} 
If gravity is described by General Relativity, the gravitational energy-momentum tensor derived from \eq{sflu} is 
\begin{equation} \label{emt}
T_{\mu\nu}=g_{\mu\nu}F-b\,F_b  (B^{-1})^{cd}\partial_\mu \Phi^c \partial_\nu \Phi^d\,,
\end{equation}
where $F_b=dF/db$. It can be identified with that of a perfect fluid defining the fluid four-velocity: 
\begin{align} \label{4vel}
u^\mu=-\frac{1}{6b\,{\sqrt{-g}}}\epsilon^{\mu\alpha\beta\gamma}\epsilon_{abc}\partial_\alpha\Phi^a\partial_\beta\Phi^b\partial_\gamma\Phi^c\,,
\end{align}
which is the solution of
\begin{align} \label{defvel}
u^\mu\partial_\mu\Phi=0\,,\quad u^2=-1.
\end{align}
This pair of conditions simply say that the fields $\Phi$ are self-transported along the fluid flow and thus define a natural notion of four-velocity. The latter can be expressed in terms of the entropy current\footnote{Notice that the entropy current is often defined with a minus sign with respect to our choice.}
\begin{align} \label{entropy}
\mathcal{J}^\mu=-b\, u^\mu\,,
\end{align}
which is covariantly and identically conserved and is just  the Hodge dual of the volume form of the fluid \cite{Dubovsky:2005xd}, as it can be seen from \eq{4vel}. Then, the quantity $b$ defined in eq.\ \eq{detb} can be interpreted as the comoving entropy density of the system:
\begin{align}
\mathcal{J}^\mu u_\mu=b\,.
\end{align}
This interpretation relies upon an appropriate identification of the temperature: $T\equiv-F_b$, leading to $b=(\rho+p)/T$, where $\rho$ and $p$ are the energy density and pressure coming from the perfect fluid energy-momentum tensor \eq{emt}. Since $\mathcal{J}^\mu$ is conserved of-shell, interpreting it as the entropy current of the fluid means that we will be dealing exclusively with non-dissipative fluids.\footnote{We will later find that the current $\mathcal{J}^\mu$ is the only object needed to construct the action for the fluid at any order in the derivative expansion.}$^{,}$\footnote{See \cite{Endlich:2012vt} for a discussion of dissipative terms in the effective field theory framework.}

The choice of the symmetry \eq{diffs} to constrain the form of the action is by no means unique. Other symmetries lead to different phenomenology for the fluid. For instance, imposing just invariance under \eq{shift} and $SO(3)$ rotations of the $\Phi$ fields, the action at lowest order in derivatives can be written as an arbitrary function of the traces of the first three powers of the matrix $B^{IJ}$ defined in \eq{Bmatrix}. This would change e.g.\ the physics of vortices in the fluid. In this work we will restrict our attention to the symmetry \eq{diffs}, which is sufficient to obtain very interesting effects for models of the acceleration of the Universe, as we will see in Section \ref{cosmos}. However, it is important to keep in mind that other choices are possible and worth exploring.

\section{Effective field theory for phonons} \label{eftph}

In this section we explore further the precise sense in which the perfect fluid action \eq{sfluid} can be interpreted as a lowest order effective field theory. This will tell us which operators have to be taken into account to study the theory at higher orders. As discussed in Section \ref{deft}, the effective theory of fluids can be constructed as a derivative expansion on the comoving coordinates $\Phi$. An object of the form $\partial\Phi$, i.e.\ a single derivative acting on one $\Phi$, is considered to be of the lowest possible order in the expansion. Hence, the leading order terms in the action are those with the same number of derivatives as $\Phi$ fields. Operators with more than one derivative per field correspond to the next-to-leading order and beyond. This is the meaning of the inequality \eq{pcphi}.  Clearly, any function of $b$, as in eq.~\eq{sfluid}, belongs in the leading order and, due to the symmetry $\mathcal{V}$Diff, $F(b)$ is the only structure allowed at that level. The next orders are then obtained from the invariants under $\mathcal{V}$Diff that are constructed with a higher number of derivatives per $\Phi$ field. 

Instead of working with $\Phi$, we can focus on $b$ and $u^\mu$ since they are invariant under \eq{diffs} and  of leading order (because they are built with just one derivative acting on each $\Phi$). This is the approach used in \cite{Bhattacharya:2012zx}, where higher order order terms are constructed counting derivatives of $b$ and $u^\mu$. 

Another possibility \cite{Leutwyler:1996er} consists in building the effective action from the fields $\pi(\tau,x)$, which are defined via
\begin{align} \label{alphapi}
\Phi(\tau,x)=x+\pi(\tau,x)
\end{align}
and represent the displacement of the fluid from a stationary state. These $\pi$ fields are often referred to as phonons, for reasons that will soon be clear. The expression \eq{sfluid} and its covariant generalization \eq{sflu} are written as actions for $\Phi$, but they can also be understood as effective field theories for the phonons. They  encode the lowest order dynamics for $\pi$ in a derivative expansion around the background \eq{ground}, given the symmetry \eq{diffs}. The equation \eq{alphapi} is just a (non-derivative) field redefinition and the $\pi$ fields are derivatively coupled (at the same order as $\Phi$) by construction. Indeed, the phonons $\pi$ are the Goldstone bosons associated to the spontaneous breaking of the internal translations \eq{shift}. 

The leading order in the phonon derivative expansion (LO) is defined by neglecting all the terms with more derivatives than $\pi$ fields. At LO each $\pi$ carries just one derivative. This allows to define a power-counting analogous to \eq{pcphi}:
\begin{align}  \label{dercountnlo}
\Lambda^{m}|\partial\pi|^n\ll|\partial^m|\partial\pi|^n|\,,
\end{align} 
where $\Lambda$ is some energy scale and $\partial$ represents a first order (space or time) derivative. In order to go beyond the LO expansion, we have to keep pieces in the $\Phi$ action that induce  higher order derivative $\pi$ terms. However, it is important to notice that we could construct the effective theory for the phonons directly in the broken phase \cite{Leutwyler:1996er}. At that level, the $\mathcal{V}$Diff symmetry would be hidden but it still determines the relations between the coefficients of the various operators in the action.

The $\pi$ fields can be used to describe small fluctuations around the background \eq{ground}. This can be applied to describe the propagation of sound waves in continuous media (hence the use of the word {\it phonon}), and for cosmological perturbation theory \cite{Endlich:2012pz,Ballesteros:2012kv,Ballesteros:2013nwa}. Then, to guarantee the validity of the expansion in the number of phonons, we assume that
\begin{equation} \label{pertlo}
|\partial\pi|\ll 1\,.
\end{equation} 
If we work purely at the phonon level, there is some ambiguity regarding which of the two conditions \eq{dercountnlo} and \eq{pertlo} should have precedence. We will assume that \eq{dercountnlo} is more stringent than \eq{pertlo}. In practical terms, this means that we must first choose the derivative order and then expand the action in $\pi$ fields.  

To obtain the effective phonon action accounting for $\mathcal{V}$Diff, we can use the set of scalar $\Phi$--operators compatible with \eq{diffs}. At leading order, there is just $b$, which leads to the action \eq{sfluid}. Then, at any order in $\pi$, the LO phonon action is obtained by expanding \eq{sfluid} in the desired number of $\pi$ fields. To this end, it is useful to write the entropy density and the four-velocity in terms of the phonon expansion around \eq{alphapi}. In Minkowski spacetime, the  entropy density is
\begin{equation} \label{detexp} 
b =1+{\partial_i\pi^i}+\frac{1}{2}\left({\partial_i\pi^i}\right)^2-\frac{1}{2}\partial_i\pi^j\partial_j\pi^i-\frac{1}{2}\left(\dot\pi^i\right)^2+\cdots\,,
\end{equation}
with the series ending with terms of six phonons. The four-velocity expansion (decomposed in time and space components) is the following:
\begin{equation} \label{fourv}
u^0 = 1+\frac{1}{2}\left(\dot\pi^i\right)^2+\cdots\,,\quad u^i = -\dot\pi^i+\dot\pi^j\partial_j\pi^i+\cdots.
\end{equation}
Neglecting metric perturbations, the LO Lagrangian for the phonons in perfect fluids is of the general form $\sum_{n}  c_n(\partial\pi)^n$ where $\partial$ represents a first order time or spatial derivative and the coefficients $c_n$ are combinations of $F$ and its derivatives evaluated on the background. Using \eq{detexp} on \eq{sfluid}, we get the quadratic LO action 
\begin{align} \label{act2}
\mathcal{S}^{(2)}[\pi]_{\text{LO}}=\frac{1}{2}(\rho+p)\int d^4x\, \left(\dot\pi_T^2+\dot\pi_L^2-c_\pi^2\left(\partial_i\pi^i\right)^2\right)\,,
\end{align}
where we have chosen to split $\pi$ into its longitudinal and transverse components:
\begin{align}
\pi=\pi_L +\pi_T\,,\quad \partial_i\pi^i_T=0\,,\quad \epsilon_{ijk}\partial_j\pi^k_L=0\,.
\end{align} 
With the notation $F_b=dF/db$, we can write as follows the energy density and pressure of the fluid 
\begin{align} \label{loeqs}
\rho=-F\,,\quad\rho+p=-b F_b
\end{align} 
and the speed of sound of longitudinal modes 
\begin{align} \label{ssound}
c_\pi^2=\frac{b {F}_{bb}}{F_b}\,.
\end{align}
The equations of motion derived from \eq{act2} are 
\begin{equation}
 \label{eomlo}
 \begin{aligned}
\ddot\pi_T^i&=0\,,\\\quad \ddot\pi_{L}^i -c_{\pi}^2\partial_i(\partial_j\pi^j_{L})&=0\,.
\end{aligned}
\end{equation}
The absence of a gradient term for the transverse modes is due to the symmetry \eq{diffs} and poses a strong coupling problem for the interpretation of the theory at the quantum level \cite{Endlich:2010hf}. 

In order to include metric perturbations from the action \eq{sflu} we need to extend our power-counting scheme, accounting for the fact that metric and matter perturbations are in general coupled.  We will consider metric fluctuations in Section \ref{cpert}, where we will study the properties of the theory  in a FLRW background.

\subsection{Fluid at NLO in derivatives} \label{nlof}

In this section we are interested in constructing the fluid effective action at next-to-leading order in derivatives (NLO). In order to obtain it, we use two basic bricks to build all the $\Phi$--operators compatible with $\mathcal{V}$Diff at NLO. One of these building blocks is $b$, the square root of the determinant of $B^{IJ}$ as defined in \eq{detb}; and the other is $u^\mu$, given in eq.~\eq{4vel}. Before doing the general analysis, we are going to look at a couple of examples for illustration \cite{Dubovsky:2011sj}.  A simple possibility with an extra derivative, and in the case of a general metric, is to consider the following piece in the Lagrangian: $\sqrt{-g}\, f(b)\, u^\mu\nabla_\mu b$. This term can be easily shown to give zero \cite{Dubovsky:2011sj}. If we define $\hat f(b)=-f(b)/b$, we can rewrite the integrand as $\sqrt{-g}\, \hat f(b)\, \mathcal{J}^\mu\nabla_\mu b$, where $\mathcal{J}^\mu$ is the entropy current \eq{entropy}. Then, defining $h(b)$ such that $\hat f=dh/db$, the integrand becomes $\sqrt{-g} \mathcal{J}^\mu\nabla_\mu h$, which gives zero after integrating by parts due to the conservation of $\mathcal{J}^\mu$. Here and in all that follows we assume that two operators differing by a total derivative are equivalent. Nevertheless, we will see that the combination $ u^\mu\nabla_\mu b$ can appear (squared) in the NLO action. 

Another operator beyond the LO that was considered in \cite{Dubovsky:2011sj}  is $y\equiv b\,\Box\, b=\partial_i\pi^i\,\Box\,\partial_j\pi^j+\cdots$. As it was discussed there, in order to include its effect on the dynamics of $\pi$, we can replace the function $F(b)$ by another function $F(b,y)$ in the fluid action. The new dependence adds an extra contribution to the phonon action \eq{act2} that
does not change the propagator for transverse modes. In fact, this operator cannot modify it at any order in the derivative expansion because \eq{detexp} does not contain a transverse linear or quadratic term. We recall that the comoving coordinates $\Phi$ have dimension $-1$ in units of energy (whereas the usual dimension for scalar fields is 1). This means that the matrix $B$ defined in \eq{Bmatrix} is dimensionless for these fields and so is $b=\sqrt{\det B}$, while $y=b\,\Box\, b$ has dimension two. Therefore, as it is always the case in effective theories, this high derivative operator must appear suppressed by some energy scale. Expanding the action $\int d^4x\, \sqrt{-g} \, F(b,y/\Lambda^2)$ to quadratic order in $\pi$ (and working in Minkowski for simplicity), the operator $y$ induces the following phonon term: $\Lambda^{-2}\int d^4x\, F_y\,\partial_i\pi^i\,\partial^2\,\partial_j\pi^j$\,, where $\partial^2=\partial_i\partial_i$, summing over the repeated indices. The size of this new term in comparison to the LO action is controlled by the small ratio $\partial^2/\Lambda^2$.  

Let us now proceed with the general construction of the action at NLO. Its form at the level of the comoving coordinates is
\begin{equation} \label{genax}
\mathcal{S}_{\text{NLO}}=\frac{1}{\Lambda^{2}}\int d^4x\, \sqrt{-g}\sum_{i}^n h_i(b) f_i\,,
\end{equation}
where the sum extends over all $n$ inequivalent operators $f_i$ built from $u^\mu$, $b$ and two derivatives. The $h_i$ are dimension-four functions of the entropy density, which must be included because $b$ is a (actually, the only) scalar  LO object.  As in the previous example with  $b\,\Box\, b$, since the $f_i$ have dimension two, we have introduced explicitly an energy scale $\Lambda$ in such a way that the $h_i$ have dimension four, like the function $F$ of \eq{sflu}. For completeness, we also require general covariance of the full action. In order to get the action for the phonons,  we simply have to expand \eq{genax} in a Taylor series around the background values of $f_i$. If we were solely interested in the phonon action, it would be equivalent to use a Lagrangian $\sqrt{-g}\tilde F\left(b\,, f_1\,,\ldots\,,  f_m\right)$ as we did before in the example with $b\,\Box\, b$. Expanding this function $\tilde F$ in $\pi$ leads to a result that can be put in one-to-one correspondence with the one from the sum of the actions \eq{sflu} and \eq{genax}. 

Among all the possible scalars that can be built with two derivatives, only six are inequivalent under integration by parts in \eq{genax}. These can be chosen to be: 
\begin{align} \label{list}
\begin{aligned}
f_1=\left(u^\mu \nabla_\mu b\right)^2\,,\quad  f_2=\nabla_\mu b \nabla^\mu b\,,\quad f_3=\nabla_\mu u^\nu \nabla_\nu u^\mu\,,\quad\quad\quad \\
\quad f_4=\epsilon_{\alpha\beta\mu\nu}\nabla^\alpha u^\beta\nabla^\mu u^\nu\,,\quad
f_5=\nabla_\mu u^\nu \nabla^\mu u_\nu\,,\quad f_6=u^\mu u^\nu \nabla_\mu u^\alpha \nabla_\nu u_\alpha\,,
\end{aligned}
\end{align}
where $\epsilon_{\alpha\beta\mu\nu}$ is the Levi-Civita symbol. Notice that we have pulled out a global $\sqrt{-g}$ factor in \eq{genax} and the operator $f_4$ only becomes a scalar if multiplied by it. A further simplification is possible using the LO equations of motion in the NLO action \eq{genax}, which amounts to performing a field redefinition \cite{Arzt:1993gz}. In this way, we can merge $f_5$ and $f_6$ into a single operator \cite{Bhattacharya:2012zx}: 
\begin{equation} \label{ost}
f_0=h^{\mu\nu}\nabla_\mu u^\alpha\nabla_\nu u_\alpha\,,
\end{equation}
where we use the projector on hypersurfaces orthogonal to the four-velocity:
\begin{align} \label{proj}
h^{\mu\nu}=g^{\mu\nu}+u^\mu u^\nu\,.
\end{align} 
Therefore, the sum in \eq{genax} runs from $i=0$ to $i=4$. These five operators agree with the ones given in \cite{Bhattacharya:2012zx}, with the exception of $f_4$, which was not taken into account there. The power-counting in derivatives of $b$ and $u^\mu$ applied in \cite{Bhattacharya:2012zx} agrees with the one that we have used. Whereas in \cite{Bhattacharya:2012zx} it was taken as an assumption motivated by the hydrodynamical approach (in which the entropy current is the fundamental object considered at the level of the equations of motion \cite{Romatschke:2009kr}), we have argued that this is the adequate power-counting for the interpretation of the construction as an effective field theory for the phonons. 

Splitting the entropy current $\mathcal{J}^\mu$ into its modulus $b$ and direction $-u^\mu$ is useful to highlight the dependence of the action on the entropy density. However, we could have arrived to the same result using $\mathcal{J}^\mu$ alone. Instead of the operators $f_0,\ldots, f_4$, we can use
\begin{align}
\begin{aligned}
 l_0=h^{\mu\nu} \nabla_\mu \J^\alpha \nabla_\nu \J^\alpha\,,\quad l_1=\left(\J^\mu\J_\alpha\nabla_\mu \J^\alpha\right)^2\,,\quad l_2=\left(\J_\alpha\nabla^\mu\J^\alpha\right)^2\,\quad \quad\\
 l_3=\nabla_\mu\J^\nu\nabla_\nu\J^\mu\,,\quad l_4=\epsilon_{\alpha\beta\mu\nu}\nabla^\alpha \J^\beta\nabla^\mu \J^\nu\,.\quad\quad\quad\quad\quad\quad
\end{aligned}
\end{align}
In terms of these, the NLO part of the action is
\begin{align} \label{jaction}
\mathcal{S}_{\text{NLO}}=\frac{1}{\Lambda^{2}}\int d^4x\,\sqrt{-g}\sum_{i=0}^4\, k_i\left(\J^2\right)\,l_i\,,
\end{align}
where the functions $k_i$ can be easily related to the $h_i$ of \eq{genax}. The action \eq{jaction} is written in terms of $\mathcal{J}^\mu$, but is still a functional of $\Phi$. It may be useful to  write this action using the current $\mathcal{J}^\mu$ as the fundamental field. The constraint $\nabla_\mu\mathcal{J}^\mu=0$ indicates that $\mathcal{J}^\mu$ has three degrees of freedom, the same as $\Phi$. We could introduce a Lagrange multiplier in \eq{jaction} to enforce this condition. Then, in order to recover $\Phi$ from $\mathcal{J}^\mu$ we would have to solve the equation $\mathcal{J}^\mu\partial_\mu\Phi=0$, which comes from eq.\ \eq{defvel}. 

Expanding the NLO action \eq{genax} at second order in $\pi$ (and working in Minkowski) we get the following contribution to the quadratic phonon action:
\begin{align} 
\mathcal{S}^{(2)}[\pi]_{\text{NLO}}=\frac{{\rho}+{p}}{2\Lambda^2} \int d^4x\, \Big( r_0 \left(\partial_i\dot\pi^j\right)^2 +r_1\left(\partial_i\dot\pi^i\right)^2+ r_2\left(\partial_j\partial_i\pi^i\right)^2+ r_3\epsilon_{ijk}\ddot\pi^i\partial_j\dot\pi^k
\Big)\,, \label{piact}
\end{align}
where $\left(\rho+p\right)r_i=2\,\tilde r_i$ and $\tilde r_0=h_0$, $\tilde r_1=h_1-h_2+h_3$, $\tilde r_2=h_2$, $\tilde r_3=-2h_4$\,. Comparing the actions \eq{piact} and \eq{act2}, it is evident that $\Lambda$ is the suppression scale for the high derivative effects on the phonon propagation. We can interpret this scale as the one at which the coarse-grained description (the fluid) needs to be replaced by a more fundamental theory. For energies and momenta sufficiently smaller than $\Lambda$, the fluid description of the system is appropriate and the level of accuracy that can be achieved in numerical calculations depends on the number of orders beyond the leading one that are taken into account. 

We can eliminate the term $\epsilon_{ijk}\ddot\pi^i\partial_j\dot\pi^k$ in \eq{piact} using the transverse linear LO equation of motion $\ddot\pi_T=0$. The argument is the same that we used before to obtain the operator $f_0$ of eq. \eq{ost}, see \cite{Arzt:1993gz}. Notice that if we had kept $f_5$ and $f_6$ instead of working with $f_0$, we would have found in addition a $\ddot\pi^2$ term that we could have subsequently eliminated as well, using again the LO (transverse and longitudinal) equations of motion for the phonons.  

We can already understand the structure of the action \eq{piact} directly at the level of the phonons. If we consider an internal $\mathcal{V}$Diff, see eq.~\eq{diffs},  $\Phi \mapsto \Psi(\Phi)$ and express these fields as $\Phi=x+\pi$ and $\Psi=x+\tilde\pi$. Then, $\Psi=\Phi+\lambda$ where $\lambda=\tilde\pi-\pi$. Taking the determinant of the Jacobian of this transformation and linearising in $\lambda$ we find that this field must be divergenceless. In addition, imposing $\dot\Psi = \dot\Phi$ we find $\dot\lambda=0$. Thus, the quadratic action has to be invariant under the transformation 
\begin{equation}
 \tilde\pi^i = \pi^i +\epsilon_{ijk} \partial_j \xi^k\,,\quad \dot \xi^i=0\,.
\end{equation} 
One can now check that this symmetry allows only the terms written in \eq{act2} and \eq{piact}. We see that eq.~\eq{diffs} prevents the phonon action $\mathcal{S}^{(2)}[\pi]_{\text{LO}}+\mathcal{S}^{(2)}[\pi]_{\text{NLO}}$ from having terms $(\partial_i\pi^j)^2$ and $(\partial_i\partial_k\pi^j)^2$. If we had only required invariance under the shifts \eq{shift} and constant $SO(3)$ rotations of $\Phi$ we would have encountered this kind of terms as well. As we have already mentioned, these two symmetries characterize what in \cite{Dubovsky:2005xd} was called a {\it solid}. The absence of transverse gradients in the quadratic phonon fluid at LO is at the root of the difficulties for making sense of the theory at the quantum level \cite{Endlich:2010hf,Gripaios:2014yha}. As we have just seen, the invariance under $\mathcal{V}$Diff implies that there are no transverse purely spatial derivatives at NLO either and thus the transverse equation of motion still allows solutions that simply evolve linearly in time. The linear equations of motion  at NLO coming from \eq{act2} and \eq{piact} can be written as: 
\begin{align}\begin{aligned}
\left(\Lambda^2 -r_0\,\partial^2\right)\ddot\pi_T^j= & 0\,,\\ 
\left(\Lambda^2-r_0\,\partial^2\right)\ddot\pi_L^j -\left(c_\pi^2 \Lambda^2+r_1\partial_0^2+r_2\,\partial^2\right)\partial_j\left(\partial_k\pi^k_L\right)= & 0\,,
\end{aligned}
\end{align}
where the sound speed $c_\pi$ and the coefficients $r_i$ would typically be of order one or smaller. Deep in the limit $\partial \ll \Lambda$ we recover the LO equations. These equations can be further simplified through another field redefinition, because the time derivatives can be completely eliminated from the NLO action \eq{piact} after integrating by parts and using the LO equations, in the same way we argued before that it can be done for the piece with $\epsilon_{ijk}$. After these manipulations, we obtain that \eq{piact} can be equivalently written as
\begin{align}
\mathcal{S}^{(2)}[\pi]_{\text{NLO}}=A\,\frac{{\rho}+{p}}{2\Lambda^2}\int d^4x\, \left(\partial_j\partial_i\pi^i\right)^2\,,\label{piact2}
\end{align}
where $A=r_2-(r_0+r_1)c_\pi^2$. This leads to the final form for the phonon equations of motion at NLO
\begin{align}\begin{aligned}
\ddot\pi_T^i=\, & 0\,,\\ 
\ddot\pi_L^i -\Big(c_\pi^2+\frac{A}{\Lambda^2}\partial^2\Big)\partial_i\left(\partial_j\pi^j_L\right)=\, & 0\,. \label{fnlo}
\end{aligned}
\end{align}
Comparing with eq.\ \eq{eomlo} we see that the NLO correction leaves unchanged the dynamics of transverse phonons  whereas it modifies the dispersion relation for longitudinal modes, with a $k^2/\Lambda^2$ suppression in momentum space characterized by the coefficient $A$. In other words, the speed of propagation of longitudinal modes becomes scale dependent. Assuming that $A$ is of order 1, the dependency of the speed on the scale is just controlled by $\Lambda$. This can be made even more explicit taking the divergence of eq.\ \eq{fnlo}  and expressing the result in terms of a field $s$ such that \begin{align}
\pi_L^i=\partial_i s\,.
\end{align}
Then, in momentum space we get: 
\begin{align}
\ddot s +\Big(c_\pi^2-\frac{A\,k^2}{\Lambda^2}\Big)k^2\, s =\, & 0\,. \label{long}
\end{align} 
 
\section{Covariant action and FLRW cosmology} \label{cosmos}

The dynamics of an effective fluid at LO in a spatially flat FLRW background was studied in \cite{Ballesteros:2012kv}. In this section we are going to focus on some of the consequences of the NLO corrections.  In order to apply the effective theory of fluids to cosmology, the first question that we need to address is how to treat gravity. A straightforward possibility consists in regarding the fluid as a matter component obeying standard gravity and hence write
\begin{align} \label{grav}
\mathcal{S}=\int d^4x\, \sqrt{-g}\left(\frac{M_P^2}{2}R  +  F(b)+
\sum_{i=0}^4 h_i(b)\frac{f_i}{\Lambda^2}\right)\,,
\end{align}
where the  $f_i$ are defined in \eq{list} and \eq{ost}. The LO action studied in \cite{Ballesteros:2012kv} corresponds to setting all the $h_i$ to zero (or equivalently to the limit $\Lambda\rightarrow\infty$).

In ref.\ \cite{Bhattacharya:2012zx}, the curvature scalar $R$ was multiplied by an arbitrary function $h(b)$, but we can show that it is actually possible to choose $h(b)=1$ without loss of generality. To see this, consider the integral $\mathcal{I} = \int d^4x \sqrt{-\tilde g}\, h(\tilde b) \tilde R$,  where $h$ is an arbitrary function and the objects carrying a tilde are built with a metric denoted by $\tilde g_{\mu\nu}$. Introducing a new metric, $g_{\mu\nu}$, via a conformal transformation of $\tilde g_{\mu\nu}(x)$ (see e.g.\ \cite{0226870332}): $g_{\mu\nu}(x)=h[\tilde b(x)] \tilde g_{\mu\nu}(x)$, the determinants of the two metrics are related by $g=h^4\tilde g$ and the Ricci scalars by  $\tilde R = h(R+3\Box \ln  h -3(\nabla_\mu \ln  h)^2/2)$. Then, defining a function $l$ such that $l(b)=h(\tilde b)$ using $b^2=[h(\tilde b)]^{-3} \tilde b^2$, we see that $\mathcal{I} =\int d^4x \sqrt{- g}(R-3f_2[l_b(b)/l(b)]^2/2 )$, and so the term that depends on $b$ can be absorbed in a redefinition of the function $h_2$ that multiplies the operator $f_2$. In addition, using that $\tilde u^\mu =\sqrt{h}\, u^\mu$, one can check that the structure of the action \eq{grav} remains the same. 

One could be concerned that once a certain $\tilde g_{\mu\nu}$ has been chosen, say FLRW, the new metric $g_{\mu\nu}$ would in general appear to be of a different kind, raising the question of whether the action with $h(b)=1$ really describes the same physics as the starting one. However, we can always find a diffeomorphism $x \mapsto x'(x)$ such that $\tilde g_{\mu\nu}(x)=g'_{\mu\nu}(x')$, which makes complete the equivalence between the two formulations. An advantage of  the Einstein frame action \eq{grav} is that it allows a straightforward definition of a gravitational energy-momentum tensor for the fluid: $T_{\mu\nu} = -2/\sqrt{-g}\, \delta \mathcal{S}_m/\delta g^{\mu\nu}$, where $\mathcal{S}_m$ is the ``matter'' part of the action  (i.e.\ the expression \eq{grav} minus the Einstein-Hilbert term). 

A conformal transformation would change the coupling between the effective fluid and the rest of the matter that may be present. If the total action (including fluid and matter fields) contains a piece of the form $\mathcal{S}_M=\int d^4x\, \sqrt{-\tilde g}\, \mathcal{L}_M[\tilde{g}_{\mu\nu},\psi_M]$, where $\psi_M$ denotes the matter fields, it will be expressed as $\mathcal{S}_M=\int d^4x\, \sqrt{-g}\, h^{-2}\mathcal{L}_M[h^{-1}{g}_{\mu\nu},\psi_M]$ after the conformal transformation $g_{\mu\nu}=h\, \tilde g_{\mu\nu}$. Then, the energy momentum tensor of the fields $\psi_M$ would become $T^{\mu\nu}[\psi_M]=h^{-1} \tilde T^{\mu\nu}[\psi_M]$, where, by definition, $\tilde T_{\mu\nu}[\psi_M] = -2/\sqrt{-\tilde g}\, \delta \mathcal{S}_M/\delta \tilde g^{\mu\nu}$. In general, only the total energy-momentum tensor of the fluid plus the matter fields will be covariantly conserved. In the effective field theory context, the most appropriate way of dealing with matter fields $\psi_M$ consists in writing their action  as $\mathcal{S}_M=\int d^4x\, \sqrt{-g}\, M(b)\mathcal{L}_M[U(b){g}_{\mu\nu},\psi_M]$, where $M$ and $U$ arbitrary functions of $b$, defined in equation \eq{detb}. Then, the effect of a conformal transformation would just be a redefinition of these functions $M$ and $U$. From the effective field theory perspective, these couplings between the fluid and the external matter fields are allowed (and should be included) in the action because, as we have argued, $b$ is an object of the lowest possible order in our power-counting scheme. Taking this into account, the previous argument for eliminating the coupling between the scalar curvature $R$ and the effective fluid extends to the general case where other matter fields are considered.\footnote{See e.g. \cite{Flanagan:2004bz, Catena:2006bd, Deruelle:2010ht, Chiba:2013mha} for related discussions in the context of scalar-tensor theories; the bottom line being that physical predictions cannot depend on field redefinitions.}

Naively, if the only role of the metric were defining the background geometry in which the fluid (and its perturbations) evolve, the action \eq{grav} would represent the simplest choice to study the effects of the NLO corrections of the fluid in cosmology. However, the metric perturbations couple to the phonons in general  (see \cite{Endlich:2012pz,Ballesteros:2012kv,Ballesteros:2013nwa}) and the gauge freedom implies that both types of perturbations should be treated within a unified and consistent power counting scheme. A possible simplifying assumption to deal with this problem is to consider that $\delta g_{\mu\nu}$ and $\partial\pi$, where $\partial$ represents any (space or time) first derivative, should be treated on an equal footing on the derivative counting. Therefore, since at NLO (and e.g.\ at quadratic order in the phonons) the action \eq{grav} generates objects such as $(\partial_i\pi^i)\partial^2(\partial_i\pi^i)$, we should also consider all the possible operators that generate terms with up to one derivative per metric perturbation.  This means that we have to include in  \eq{grav} all the operators with up to two derivatives of the metric. Since the Riemann tensor contains already two derivatives exactly, there are just two such terms \cite{Bhattacharya:2012zx}. One of them is the scalar curvature $R$, which we already take into account in the Einstein-Hilbert piece of the action \eq{grav}. The other one is $R_{\mu\nu}u^\mu u^\nu$, which is redundant because it can be expressed in terms of the $f_i$ operators after integration by parts \cite{Bhattacharya:2012zx}. The action \eq{grav} is the simplest one that allows to study the NLO effects of the fluid once gravity is taken into account. If we wanted to go to higher orders in derivatives we would have to consider higher order contractions of Riemann tensors and four-velocities.\footnote{For a review of gravity as an effective theory see \cite{Burgess:2003jk}.} 

\subsection{Background evolution} \label{bkg}
We will denote the FLRW scale factor by $a(\tau)$ and the conformal Hubble parameter by $\ch=\dot a/ a$, where $\dot a= d a/d\tau$.
The background evolution derived from \eq{grav} is governed by the equation
\begin{align} \label{fried}
3M_P^2\left(1+\frac{m^2}{\Lambda^2}\right)\ch^2+a^2 F=0\,,
\end{align}
where $m^2$ is a function of $b=a^{-3}$ with units of mass squared (that can be of either sign):
\begin{align} \label{m}
m^2 M_P^2=-h_0-3\,b^2\,(h_1-h_2)-h_3\,.
\end{align}
The eq.\ \eq{fried} is just the Friedmann equation for the energy density
\begin{align} \label{fr1}
3M_P^2\ch^2=a^2\rho\,.
\end{align} 
In the limit of very large $\Lambda$, we recover the background energy density of the LO fluid, which is $\rho=-F$. It could then be tempting to interpret eq.\ \eq{fried} as the background evolution of the LO fluid in a modification of gravity given by a varying Planck mass:
\begin{align}\label{meff}
M_{\text{eff}}^2=M_P^2\left(1+\frac{m^2}{\Lambda^2}\right)\,.
\end{align}
From eqs.\ \eq{grav}, \eq{fried} and \eq{m} it is natural to expect that $m^2\sim h_i/M_P^2\sim F/M_P^2\sim\ch^2$, which means that the NLO corrections to the Planck mass are of $\mathcal{O}\left({\ch^2}/{\Lambda^2}\right)$. This is basically  a consequence of assuming that the free functions of $b$ that appear in the action \eq{grav} are all of the same order. The ratio $\ch^2/\Lambda^2$ will also appear later as the parameter that controls the size of the corrections to the equations of the perturbations. 

The equation of state of the effective fluid at LO is $w=-1+b F_b/F$ (see e.g.\ \cite{Ballesteros:2012kv}) and the choice $F\propto b$ corresponds to cold dark matter ($w=0$). We can then easily estimate the size of $m^2$ that would be required to obtain late time acceleration (in agreement with the data) from the NLO correction. If we define LO and NLO density parameters as  $\Omega_{\text{LO}}= -F/(3M_P^2\ch^2)$ and $\Omega_{\text{NLO}}=-m^2/(a^2\Lambda^2)$, the condition that today $\Omega_{\text{NLO}}\sim\Omega_{\text{LO}}\sim 1/2$  gives $m^2/\Lambda^2\sim 1$, which implies an order 1 correction to the Planck mass (see eq.\ \eq{meff}). If we assume that $h_i \sim F$ (which is the normal expectation given the action \eq{grav}) and $m\sim \Lambda$, then $\Lambda$ has to be of the order of $\ch$, which would be in conflict with the natural regime of validity of the effective theory.  Therefore, even if interpreting only the NLO corrections as a modification of gravity can be appealing, it would seem to be unnatural to use these corrections alone to describe self-acceleration at late times. This result is consistent with our choice of power-counting scheme for the metric perturbations and with our previous estimates in Section \ref{deft}. Given that $R\sim \Lambda^4/M_P^2\ll\Lambda^2$ for $\Lambda \ll M_P$ and, in addition, $\ch^2\sim R$\,; if as we argued before $m^2\sim \ch^2$, then $m^2\sim\Lambda^2$ would only be possible if $\Lambda\sim M_P$ which would contradict our initial assumption that $\Lambda\ll M_P$ and imply that higher powers of the curvature $R$ should not be neglected.

There exists no impediment however in applying the whole LO+NLO action \eq{grav} to build models of dark energy.  As we will see in the next section, this approach leads to interesting phenomenology for cosmological perturbations, with concrete predictions that make the theory distinguishable in principle from a cosmological constant and other types of models of the acceleration. 

The equation of state ($p=w\rho$) at NLO order can be expressed as
\begin{align} \label{sstate}
w= \frac{a^2F}{3M_P^2\ch^2}\left(1+\frac{1}{3}\frac{ d\ln F}{d\ln a}\right)+\frac{m^2}{3\Lambda^2}\left(1+\frac{d\ln\left(m^2\ch^2\right)}{d\ln a}\right)\,,
\end{align}
which reduces to the LO expression for $\Lambda\rightarrow \infty$. Since $m^2/\Lambda^2$ is small compared to 1, in order to obtain $w\sim -1$ the function $F$ should be chosen such that $d \ln F/ d \ln a\ll 1$, which is equivalent to the condition $b\,F_b\ll F$. 

We can write the pressure using the time derivative of eq.\ \eq{fried}, which leads to 
\begin{align} \label{fr2}
6M_P^2\dot\ch+a^2(\rho+3p)=0\,,
\end{align} 
and shows that the two Friedmann equations, \eq{fr1} and \eq{fr2}, are not independent. As usual, it follows from them that 
\begin{align}
\dot\rho+3\ch(\rho+p)=0\,,
\end{align}
and therefore the equation of state and the adiabatic sound speed 
\begin{align}
{\dot p}\equiv c_A^2{\dot \rho}
\end{align} 
are related by 
\begin{align}
\dot w = 3\ch(1+w)\left(w-c_A^2\right).
\end{align}

\subsection{Cosmological perturbations} \label{cpert}

In this section we focus on linear perturbations at NLO in a flat FLRW Universe in Poisson gauge. We use the notation of reference \cite{Ballesteros:2012kv} and write the perturbations of the metric as follows: $\delta g_{00}=-2a^2\psi$, $\delta g_{0i}= a^2 \nu^i$ and $\delta g_{ij}= a^2\left[(1-2\phi)\delta_{ij}+\chi_{ij}\right]$, where 
$\partial_k \nu^k=\partial_k\chi_{kj}=0$ and $\chi_{jj}=0$. As we have just seen, the NLO modification of the background evolution is given by a single combination, $m^2$, of the functions $h_i$. The corrections to the energy density and pressure perturbations of the fluid, $\delta \rho$, and $\delta p$ are  determined by $m^2$ and another function $\gamma$, see the Appendix and eq.\ \eq{threef}. Other combinations of the $h_i$ appear in the full description of linear perturbations. The propagation of gravitational waves and the (scalar) anisotropic stress (which is zero at LO) are affected by the same function, that we will denote by $\alpha$. The vector perturbations of the metric, which couple only to transverse phonons, are characterized by $\alpha$ and three different combinations of the $h_i$ functions: $\beta$, $\gamma$ and $\xi$. These are the only modes affected by the operator $\sqrt{-g}f_4$. We will now study each type of perturbation in turn.

\subsubsection{Gravitational waves}

The propagation of tensor modes at NLO is given by the quadratic action 
\begin{align} \label{wavesaction}
\mathcal{S}^{(2)}[\chi]=\frac{M_P^2}{8}\int d^4x\, a^2\left(\alpha\left(\dot\chi_{ij}\right)^2-\left(\partial_k\chi_{ij}\right)^2\right)\,,
\end{align}
where the time dependent function
\begin{align} \label{alphag}
\alpha=1+\frac{2(h_0+h_3)}{\Lambda^2 M_P^2}
\end{align}
determines the speed of propagation of the gravitational waves. The action \eq{wavesaction} has the same form as the one that appears in the effective field theory of inflation \cite{Creminelli:2006xe,Cheung:2007st} from a quadratic term in the perturbation of the extrinsic curvature of spatial slices, see e.g. \cite{Creminelli:2014wna}.\footnote{See also \cite{Cannone:2014uqa} for a generic form of the action for tensor perturbations allowing for the breaking of spatial diffeomorphisms. In that case, a mass term for the fluctuation $\chi_{ij}$ appears. This term is not present in \eq{wavesaction} due to the symmetry $\mathcal{V}$Diff.}   Since tensor modes are gauge invariant, the action \eq{wavesaction} is the same for any other choice of	 gauge. 

In standard gravity, the speed of propagation of tensor modes is equal to the speed of light and the presence of an effective fluid at LO does not change this result \cite{Ballesteros:2012kv}. However, the NLO fluid operators induce a correction to the quadratic action of order $\sim\ch^2/\Lambda^2$. The propagation equation resulting from \eq{wavesaction} is
\begin{align} \label{waves}
\ddot\chi_{ij}+\left(2+\frac{d \ln \alpha}{d \ln a} \right)\ch\dot\chi_{ij}-\frac{\partial^2\chi_{ij}}{\alpha}=0\,.
\end{align}
The (time-varying) speed of propagation of gravitational waves can then be defined as
\begin{align}
c_\chi^2=\frac{1}{\alpha}=1-\frac{2(h_0+h_3)}{\Lambda^2 M_P^2}+\mathcal{O}\left(\frac{\ch^4}{\Lambda^4}\right)\,.
\end{align}
For $\Lambda\gg\ch$ we recover the standard result $c_\chi=1$. In order to get an estimate of the modification to the  Hubble friction term in \eq{waves}  it is reasonable to assume that $dh_i/db\sim dF/db$, which implies $d \ln \alpha/d \ln a\sim 1$. 

Models of a single derivatively scalar with second order equations of motion can also modify the friction coefficient and the speed of propagation of gravitational waves. A broad class  of them, called {\it Horndeski models}, was first constructed in \cite{Horndeski:1974}. The structure of the action for gravitational waves in those models (see e.g. \cite{Bellini:2014fua}) is similar to our equation \eq{wavesaction}. However, in that case it is not possible to estimate the size of $\alpha$ and its derivatives from the structure of the theory itself, as we have done here. 

Following \cite{Bellini:2014fua}, we can define (yet another) effective Planck Mass, $M_*$, from the coefficient of the action giving the proper normalization of the kinetic term of the graviton.\footnote{This $M_*$ should not be confused with the quantity $M_{\text{eff}}$ defined in \eq{meff}, which has a different physical meaning.} As shown by equation \eq{waves}, this definition is motivated by the fact that gravitational waves can propagate in absence of matter sources. Reading from \eq{wavesaction}, we see that in our case 
\begin{align}
M_*^2 = \alpha\, M_P^2\,,
\end{align}
and therefore $M_*$ is completely given at NLO by the speed of propagation of gravitational waves $c_\chi^2=1/\alpha$. In general, in the class of models of \cite{Horndeski:1974}, these two quantities are independent from each other \cite{Bellini:2014fua}. We expect that $M_*^2$ and $c_\chi^2$ will also become independent in the effective theory of fluids by including higher order terms in the derivative expansion.  

The idea of measuring $c_\chi$ and the friction excess $d \ln \alpha/d \ln a$ (at early times, but after inflation) using the B-mode polarization of the CMB has been recently explored   in \cite{Amendola:2014wma,Raveri:2014eea,Pettorino:2014bka}. In general, a modified dispersion relation for gravitational waves may be used as a signature for deviations from $\Lambda$CDM.

\subsubsection{Transverse phonons and vector modes}

In ref.\ \cite{Ballesteros:2012kv} it was shown that the conservation of vorticity (which is due to the invariance under $\mathcal V$Diff) determines the dynamics of transverse modes and vector metric perturbations at leading order in derivatives. When the NLO contributions are taken into account, their evolution is modified by terms of order $\ch^2/\Lambda^2$ that are controlled by $\alpha$ of eq.\ \eq{alphag} and three more functions of time:
\begin{align} \label{threef}
\beta=1+\frac{2(h_0-h_3)}{\Lambda^2M_P^2}\,,\quad \gamma = 1+\frac{6\,b^2h_2}{\Lambda^2 M_P^2}\,,\quad \xi = \frac{4\,h_4}{\Lambda^2 M_P^2} \,.
\end{align}
The conservation equation that gives the time evolution of transverse phonons is: 
\begin{align} \label{cons}
\dot{Q}^i=0\,,
\end{align}
where 
\begin{align}
Q^i=3\ch^2 a^2 (\gamma+w)\dot\varphi_T^i+a^2(1-\alpha)\partial^2\left(\nu^i+\dot\varphi^i_T\right)-\frac{\epsilon_{ijk}}{a^2}\partial_j\left(2\,\xi\,\ch\dot\varphi_T^k-\frac{\dot\xi}{2}\left(\dot\varphi_T^k+\nu^k\right)\right)\,,
\end{align}
and we have defined the combination
 \begin{align} \label{git}
\dot\varphi^i_T=\dot\pi^i_T-\nu^i\,,
\end{align} 
which is related to the three-momentum of the fluid in the LO approximation \cite{Ballesteros:2012kv}. The interpretation of \eq{cons} as a consequence of the conservation of vorticity was discussed in \cite{Ballesteros:2012kv} in the LO case and FLRW. For the construction of the vorticity charges in Minkowski at LO see \cite{Dubovsky:2005xd} and \cite{Ballesteros:2012kv}. 

The Einstein equation for the vector perturbations of the metric is
\begin{align} \label{E2}
\beta\, a^2\partial^2 \nu^i+6\,a^2\ch^2(\gamma+w)\dot\varphi_T^i-\frac{\epsilon_{ijk}}{a^2}\partial_j\left(4\,\ch\,\xi\,\dot\varphi_T^k+\dot\xi\, \nu^k\right)=0\,.
\end{align}
The usual equations for a generic phenomenological fluid in General Relativity in FLRW are recovered from \eq{cons} and \eq{E2} in the limit $\xi=0$ and $\alpha = \beta = \gamma = 1$\,, which corresponds to the LO action of the effective fluid. In standard gravity, vector perturbations decay in an expanding universe. This is the reason why they are often neglected, e.g.\ in studies of inflation. We see that the effective field theory of fluids at NLO features deviations from the usual behaviour of order $\ch^2/\Lambda^2$. This does not generally occur in models of a single scalar coupled to gravity.

\subsubsection{Longitudinal phonons and scalar modes} \label{longpho}

A phenomenological ``sound speed'' that relates pressure and density perturbations
\begin{align}
\delta p = c_s^2 \delta \rho
\end{align} is often introduced in cosmology. Defined in this way, $c_s^2$ plays for the perturbations a similar role to that of the equation of state ($w=p/\rho$) for the background.  If $c_s^2=c_A^2$, where $c_A^2=\dot p/\dot \rho$, the fluid is said to be adiabatic. This is the case e.g.\ for standard quintessence with a canonical kinetic term and a potential.

At leading order in derivatives, the speed of propagation of longitudinal modes in an effective fluid in FLRW is time dependent and can be computed using the expression \eq{ssound}, which is the same as in Minkowski \cite{Ballesteros:2012kv}. It was also shown in \cite{Ballesteros:2012kv} that \eq{ssound} is equal to the adiabatic speed of sound $c_A^2$ and to $c_s^2$ at LO, which makes the fluid adiabatic at that order. However, we found at the end of Section \ref{nlof}  that in Minkowski space the NLO corrections induce a scale dependence  in the speed of longitudinal modes $c_\pi^2$ of order $k^2/\Lambda^2$, see eq.\ \eq{fnlo}. In FLRW this translates into the breaking of the relation $c_s^2=c_A^2$, with corrections suppressed by $\ch^2/\Lambda^2$. The equations \eq{deltar} and \eq{deltap} of the Appendix are the NLO expressions for $\delta \rho$ and $\delta p$ at linear order in perturbations, showing explicitly these corrections. Their sizes are controlled by the function $m^2$ that enters in the background evolution and by the function $\gamma$, see \eq{threef},  that comes from the operator $f_2=\nabla_\mu b \nabla^\mu b$ defined in \eq{list}.

The other important feature that appears at NLO is an anisotropic stress component $\sigma$:
\begin{align}
(\rho+p)\partial^2\sigma=\left(\partial^{i}\partial ^{j}-\delta^{ij}\frac{\partial^2}{3}\right)T_{ij}\,.
\end{align}
At leading order in derivatives in the fluid effective action \eq{grav} (i.e.\ keeping only the $F(b)$ piece) we find that  $\partial^2\sigma =0$ \cite{Ballesteros:2012kv}. The NLO contributions to $\sigma$ is
\begin{align}\label{anist}
12\pi G\,(\rho+p)\,\partial^2\sigma=(-\dot\alpha+2(1-\alpha)\ch)\partial^2\partial_i\dot\pi^i+(1-\alpha)\partial^2\partial_i\ddot\pi^i\,.
\end{align}
This equation can be interpreted as an equation of state for the anisotropic stress $\sigma$ in terms of the velocity divergence $\theta=-\partial_i\dot\pi^i$ and its first time derivative, see also eqs.\ \eq{spatv} and \eq{theta} of the Appendix. The anisotropic stress depends on $\alpha$, introduced in eq.\ \eq{alphag}, which is the same function of time that modifies the propagation of gravitational waves at NLO.  A similar connection between the anisotropic stress and the speed of propagation of gravitational waves in single-scalar models was exploited in \cite{Amendola:2014wma} to study the effects of modified gravity on CMB B-modes.

The possible presence of anisotropic stress at late times in the history of the Universe has often been regarded in the literature as a ``smoking gun'' for the detection of deviations from $\Lambda$CDM (and in turn as a signature of modified gravity/dark energy). The anisotropic stress affects the relation between the metric potentials, eq.\ \eq{metricpot},  and physical phenomena such as e.g.\ weak gravitational lensing of the CMB. As discussed in Section \ref{bkg}, an attractive picture in the case of effective fluids is to consider the NLO corrections as a mild modification of gravity (controlled by the scale $\Lambda$) in which the LO fluid evolves. A more powerful application consists in using the entire LO+NLO$+\cdots$ action to construct models of modified gravity/dark energy (characterized by the functions $F$, $h_i$, ...) in which the components of the universe (e.g.\ dark matter) propagate. In both cases the anisotropic stress is naturally small since it is determined by corrections of the size of $\ch^2/\Lambda^2$. This makes the effective theory of fluids an interesting playground for models of the acceleration with dynamics close to $\Lambda$CDM and yet different from it, with potentially observable consequences. 

\section{Conclusions and Outlook} \label{conc}
We know very little about the actual nature of dark energy and the current data are fully compatible with the acceleration of the Universe being due to a small cosmological constant (see e.g.\ \cite{Ade:2013zuv}). In order to search for alternatives beyond the standard $\Lambda$CDM model, we should aim for general, well-motivated and consistent ways of describing the acceleration that allow to draw concrete testable predictions. For several years we have entertained the possibility that the acceleration could be due to a modification of Einstein's gravity at very large distances or to an exotic component of the Universe with negative pressure. The divide between these two interpretations is blurry, but it is clear that both of them require extra degrees of freedom. A powerful approach in this context can be built from the idea that,  below some energy scale, these new degrees of freedom form a fluid whose properties are determined by internal symmetries. The framework of effective field theories is the natural playground for implementing this picture.

In this paper we have explored the effective theory of fluids at next-to-leading order in derivatives and we have applied it in a cosmological context, showing  that it leads to very interesting phenomenology for describing the acceleration of the Universe, comparable to that of popular single-field scalar models. The signatures of the theory at NLO include anisotropic stress, a non-adiabatic speed of sound of scalar modes, modifications to the dynamics of vortices (vector modes) and also to the propagation of gravitational waves. This richness and the fact that the size of these effects is controlled by an energy scale make the theory particularly attractive for modelling the acceleration beyond a cosmological constant. In addition, the effective framework guarantees that high order derivative corrections are kept under control without having to resort to any ad-hoc fine tuning between different operators. 

Bearing in mind the assumptions that we have made, in particular about the symmetries and the field content of the fluid, a detection of a $\Lambda$CDM deviation of the kinds predicted by the theory (e.g.\ anisotropic stress) can be used to estimate within this framework the energy scale $\Lambda$ responsible for the NLO corrections.\footnote{Just to illustrate the point, and setting aside for a moment the obvious and large qualitative differences, an analogy in particle physics  would be using an LHC detection of an anomalous cross-section to infer physics from operators beyond the Standard Model.}

We have restricted our analysis to the rather minimal situation of a fluid with invariance under volume preserving diffeomorphisms. In this case, there are just three new degrees of freedom, which appear naturally, each corresponding to one space dimension. For perturbations in a FLRW Universe, this implies that there are two (transverse) fields that couple to metric  vector modes and one (longitudinal) that interacts with scalar fluctuations. In total, six operators of the entropy current $\mathcal{J}^\mu$ of eq.\ \eq{entropy},  see eqs.\ \eq{grav}, \eq{list} and \eq{ost}, are added to the Einstein-Hilbert action up to NLO in derivatives. 

At the level of the background evolution, the intrinsic NLO terms modify the Friedmann equations by a single combination $m^2$ of the functions of the entropy density $b=u^\mu \mathcal{J}_\mu$ multiplying these operators. Four more independent functions:\ $\alpha$, $\beta$, $\gamma$ and $\xi$, appear from these operators at the level of linear cosmological perturbations. All of these functions intervene on the dynamics of vector modes, whereas the function $\alpha$ is the only one responsible for  the anisotropic stress and the modification to the propagation of gravitational waves. The same combination, $m^2$, that enters in the background expansion appears also in the energy density and pressure perturbations. The latter is also affected by one of the functions, $\gamma$,  that determines the dynamics of vector modes. As we have emphasized, the size of the NLO corrections is controlled by a scale, which determines the regime of validity of the theory. This ensures that the terms with high order derivatives in the equations of motion do not develop instabilities within that regime.

The equations presented here may be applied not only in the context of dark energy but also to other cosmological species. For example, we could use this framework (including the NLO corrections) to describe baryons and dark matter as imperfect fluids. In fact, in a general treatment, all the matter components may in principle be described using the formalism of effective fluids, even if there are non-gravitational interactions among the species. This has been discussed at LO in derivatives in \cite{Ballesteros:2013nwa}. 

The idea of using a field theory for describing matter species in cosmology has been applied before. For instance, a {\it k-essence} model \cite{ArmendarizPicon:1999rj,Garriga:1999vw,ArmendarizPicon:2000dh} leads to a perfect barotropic energy-momentum tensor and was used in \cite{Boubekeur:2008kn} to describe the period of matter domination after inflation. Notice that such a system is necessarily irrotational, and the same occurs in the case of Hordenski models \cite{Horndeski:1974}. This is different from the effective theory of fluids, in which two additional degrees of freedom allow the existence of transverse modes. 

Following this idea, the application of our framework to dark matter can allow to describe some effects at large scales that go beyond the standard cold dark matter picture. At LO, cold dark matter corresponds to the choice $F(b)\propto b$, see eqs.\ \eq{loeqs} and \eq{ssound}. Once the NLO corrections are taken into account, an appropriate choice of $F$ and $m^2$ ensures that the condition $w\simeq0$ can be maintained, see eq. \eq{sstate}. Then, using the NLO terms we can simultaneously obtain a small anisotropic stress and a speed of sound different from zero, making possible the construction of a generalized \cite{Hu:1998kj} or warm dark matter model. As in the usual cold dark matter scenario, the vorticity will typically decay during matter domination, albeit with scale dependent corrections modulated by the size of $\Lambda$. In this context, let us also notice that the predictions for the propagation speed and the friction term of gravitational waves are general at NLO and do not depend on the assumption of a dark energy component.

Other symmetry choices different from $\mathcal{V}$Diff should also be explored and novel effects can be expected to occur. For example, extending these NLO results to solids (invariant under internal translations and $SO(3)$ rotations) will lead to different properties for transverse modes and gravitational waves. This may also have interesting implications for models of inflation such as {\it solid inflation} \cite{Gruzinov:2004ty, Endlich:2012pz}. In this respect, a systematic analysis using the CCWZ approach \cite{Callan:1969sn}, as it has already been applied at LO \cite{Nicolis:2013lma}, is worth considering. Extending the theory presented here to include extra conserved charges or even supersymmetry are other directions for the future. Finally, the study of the NLO corrections in Minkowski might also be useful for understanding the quantum nature of effective fluids \cite{Endlich:2010hf,Gripaios:2014yha}.

\subsubsection*{Acknowledgments}
I thank B.\ Bellazzini, E.\ Bellini, J.\ A.\ Casas,  J.\ R.\ Espinosa, J.\ Garriga, C.\ Hoyos, L.\ Mercolli and C.\ Tamarit for very valuable discussions and/or comments on a draft of this paper. Some of the computations at NLO have been done with the Mathematica packages xAct (\url{www.xact.es}) and xPand \cite{Pitrou:2013hga}. I thank the authors of the last, E.\ Bellini and J.\ Sakstein for conversations concerning these packages. This research was supported by the ``TRR33 -- The Dark Universe'' DFG Grant. I thank the hospitality of Perimeter Institute for Theoretical Physics while part of this work was done. [Research at Perimeter Institute is supported by the Government of Canada through Industry Canada and by the Province of Ontario through the Ministry of Economic Development \& Innovation.]

\section*{Energy-momentum tensor and gauge invariant variables}
It is useful to decompose the energy momentum tensor in projections with respect to a four-velocity frame $U^\mu$, defining the energy density $\rho_{(U)}$ and the pressure $p_{(U)}$ as
\begin{align}
\rho_{(U)} = T_{\mu\nu} U^\mu U^\nu\,,\quad 3\,p_{(U)} = T_{\mu\nu}h_{(U)}^{\mu\nu}\,,
\end{align}
where $h_{(U)}^{\mu\nu}=U^\mu U^\nu+g^{\mu\nu}$ is a projector analogous to \eq{proj}. The frame $U^\mu$ can be arbitrarily chosen, but it is often  convenient to work with the four-velocity corresponding to the energy (or rest) frame of the fluid; such that the energy flux
\begin{align} \label{heat}
q_{(U)}^\mu = - h_{(U)}^{\nu\mu}T_{\nu\gamma}U^{\gamma}\,,
\end{align}
vanishes. This is the most common choice in cosmological perturbation theory (see e.g.\ \cite{Bardeen:1980kt}). In our case however, the four velocity $u^\mu$ introduced in \eq{defvel} does not give a zero energy flux at NLO. The expansion of $u^\mu$ in a perturbed FLRW universe in Poisson gauge is \cite{Ballesteros:2012kv} 
\begin{equation}
\begin{aligned} \label{fourvg}
u^0 =\frac{1}{a}\left(1-\psi+\nu^iv^i+\frac{3}{2}\psi^2+\frac{1}{2}v^2+\cdots\right)\,,\quad u^i = \frac{1}{a}\left(1-\psi+\cdots\right)v^i\,,
\end{aligned}
\end{equation}
where
\begin{align} \label{spatv}
v^i=\frac{d x^i}{d \tau} =  -\dot\pi^i+\dot\pi^j\partial_j\pi^i+\cdots.
\end{align}
is equal to the spatial part of the four-velocity in Minkowski \eq{fourv}. Choosing $U^\mu=u^\mu$, the spatial part of the projection \eq{heat} is
\begin{align}
\label{heats}
\frac{q^i}{a\, M_P^2}= \frac{1-\beta}{2a^2}\partial^2 \nu^i+3\,\frac{1-\gamma}{a^2}\ch^2\left(\dot\varphi^i-\partial_i\hat\pi\right)-\frac{1}{2}b^2\dot \xi\, \epsilon_{ijk}\,\partial_j\dot\varphi^k
\end{align}
where $\dot\varphi^i =\dot\pi^i-\nu^i$ is the analogous of the field  \eq{git} (but now including the longitudinal part) and 
\begin{align}
\hat\pi=-\frac{1}{3\ch}\left(\partial_i\pi^i+3\phi\right)\,,
\end{align}
can be identified with the time shift that allows to define adiabatic modes at LO \cite{Ballesteros:2012kv}. 

The expression \eq{heats} only vanishes in general for the LO fluid, but not at NLO. This needs to be taken into account to make contact with the usual equations of scalar perturbations, see e.g.\ \cite{Bardeen:1980kt, Ma:1995ey}. The difference with respect to the usual case (where $q^\mu=0$) is that the divergence of the 0--$i$ Einstein eq.\ \eq{eq0i} contains a term that depends on $\Xi$, which is the divergence of the spatial part of the energy flux: 
\begin{align}
a\,(\rho+p)\,\Xi=\partial_i q^i\,.
\end{align}
This piece can be absorbed in a redefinition of the velocity divergence by changing to a frame $\mathcal{U}^\mu$ with $q^\mu_{(\mathcal U)}=0$. At linear order in perturbations, a change of frame does not affect the energy density, the pressure and the anisotropic stress; see e.g. \cite{Clarkson:2010uz}, and therefore the other three Einstein equations for scalar perturbations are unaffected. The four of them at NLO are:
\begin{align}
3\ch\dot\phi+3\ch^2\psi-\partial^2\phi=-4\pi G a^2 \delta \rho\\ \label{eq0i}
\partial^2\left(\dot\phi+\ch\psi\right)=-4\pi G a^2 (\rho+p) \left(\theta-\Xi\right)\\
\ddot\phi+\ch(\dot\psi+2\dot\phi)+\left(\ch^2+2\dot\ch\right)\psi+\frac{\partial^2}{3}\left(\psi-\phi\right)=4\pi G a^2 \delta p\\ \label{metricpot}
\partial^4(\psi-\phi)=12\pi G a^2 (\rho+p)\partial^2\sigma\,,
\end{align}
where
\begin{align} \label{theta}
\theta=-\partial_i\dot\pi^i
\end{align}
is the linear part of the divergence of the spatial velocity \eq{spatv}\,. The anisotropic stress $\partial^2\sigma$ was given in eq.\ \eq{anist} and the linear perturbations of the energy density and the pressure are
\begin{align}  \label{deltar}
&\delta\rho = -3\ch\left(1+\frac{m^2}{\Lambda^2}\right)(\rho+p)\hat\pi-6M_P^2\frac{m^2}{a^2\Lambda^2}\ch\left(\left(\ch \hat\pi\right)^{\LargerCdot}-\ch\psi\right)\,,\\ \nonumber
&\delta p = \left(-3\, c_A^2\ch \left(1+\frac{m^2}{\Lambda^2}\right)+\frac{\left(m^2\right)^{\LargerCdot}}{\Lambda^2}\right)(\rho+p)\hat\pi+2\frac{m^2}{\Lambda^2}(\rho+p)\psi+2\frac{M_P^2}{a^2\Lambda^2}\left(m^2\right)^{\LargerCdot}\left(\left(\ch \hat\pi\right)^{\LargerCdot}-\ch\psi\right)\\
& \quad +M_P^2\frac{\gamma-1}{a^2}\ch\left(\partial_i\dot\pi^i-\partial^2 \hat\pi\right)+2M_P^2\frac{m^2}{a^2\Lambda^2}\left(\left(\ch \hat\pi\right)^{\LargerCdot\LargerCdot}+2\ch\left(\ch \hat\pi\right)^{\LargerCdot}-\ch\dot\psi-3\ch^2\psi\right)\,. \label{deltap}
\end{align}
Taking the limit $\Lambda\rightarrow \infty$ we obtain the LO relation $\delta p= c_A^2\delta\rho$, see the discussion in Section \ref{longpho}. 

Defining $\theta_R=\theta-\Xi$ and differentiating \eq{eq0i} with respect to time we can write the usual Euler equation for the evolution of the velocity divergence in the rest frame:
\begin{align}
\dot\theta_R+(1-3c_A^2)\ch\theta_R+\frac{\partial^2\delta p}{\rho+p}-\partial^2\sigma+\partial^2\psi=0\,.
\end{align}
This  is nothing but the equation that describes the evolution of longitudinal phonons written in a compact way. At NLO this equation contains high derivative terms like $\partial^2 \partial_i\ddot\pi^i$, but this does not introduce any instability problem since these terms are all suppressed by $\Lambda^{-2}$. 

Although we have worked in Poisson gauge for simplicity, it is straightforward to connect the results with gauge-invariant variables. As it is well-known, a gauge transformation can be seen as a change of coordinates in the perturbed (physical) spacetime, keeping fixed the coordinates of the background (fictitious) spacetime \cite{Bardeen:1980kt}. A general gauge transformation can then be written as 
\begin{align} \label{gauge}
\tilde\tau =\tau+T(\tau,x^j)\,,\quad \tilde x^i = x^i+ L^i(\tau,x^j)\,,
\end{align}
where the spatial part can be decomposed as $L^i=\partial_i L+l^i$ with $\partial_il^i=0$. 

By definition, tensor perturbations $\chi_{ij}$ are traceless and divergenceless and therefore they are invariant under gauge transformations. 

The most general vector perturbation of the metric is given by $\delta g_{0i}=a^2\nu_i$ and $\delta g_{ij}= 2a^2\partial_{(i}h_{j)}$. Under a gauge transformation $\tilde \nu^i=\nu^i-\dot l^i$ and $\tilde h^i = h^i-l^i$. This implies that $B^i=\nu^i-\dot h^i$ is gauge invariant.  In Poisson gauge $h^i=0$ and therefore $B^i=\nu^i$ in this gauge. 

Since the fields $\Phi^i$ are scalars, they satisfy $\tilde\Phi^i(\tilde\tau,\tilde x)=\Phi^i(\tau,x)$. Expanding this equation in phonons we have $\tilde x^i+\tilde  \pi^i(\tilde\tau,\tilde x) = x^i+\pi^i(\tau,x)$ and using \eq{gauge} we get that $\tilde  \pi^i=\pi^i-L^i$ at linear order, in agreement with the linear transformation rule for the spatial velocity $v^i$ of eq.\ \eq{spatv}. Therefore, the combination $\dot\varphi^i=\dot\pi^i_T-\nu^i$ is gauge invariant. In Poisson gauge this corresponds to the field that we defined in equation \eq{git}.

The longitudinal phonons transform according to $\tilde\pi_L^i=\pi_L^i-\partial_i L$ and so $\tilde\partial_i\tilde\pi^i=\partial_i\pi^i-\partial^2L$. If we write $\delta g_{ij} = 2a^2 \partial_i\partial_j E$, we obtain that $E$ transforms according to $\tilde E=E-L$ and therefore $\Upsilon = \partial_i\pi^i-\partial^2 E$ is gauge invariant. In Poisson gauge $E=0$ and $\partial_i\pi^i$ is proportional at LO to the comoving curvature perturbation on uniform density hypersurfaces \cite{Ballesteros:2012kv}. Besides, the gauge invariant Bardeen potentials $\Phi$ and $\Psi$  \cite{Bardeen:1980kt} reduce to the metric perturbations $\phi$ and $\psi$ in Poisson gauge.

\bibliographystyle{hunsrt}
\bibliography{fluidbib}

\end{document}